\documentclass[twocolumn,showkeys,showpacs,preprintnumbers,prd,superscriptaddress,nofootinbib]{revtex4-1}

\usepackage{graphicx}
\usepackage{epsf}
\usepackage{bm}
\usepackage{amsmath}
\usepackage{amsfonts}
\usepackage{amssymb}
\usepackage{epstopdf}
\usepackage{natbib}
\usepackage{hyperref}
\usepackage{color}
\usepackage{verbatim}
\usepackage{multirow}
\usepackage{bm}
\usepackage{hyperref}
\begin{document}

\title{Weighing massive neutron star with screening gravity: A look on PSR J0740+6620 and GW190814 secondary component}

\author{Rafael C. Nunes}
\email{rafadcnunes@gmail.com}
\affiliation{Divis\~ao de Astrof\'isica, Instituto Nacional de Pesquisas Espaciais, Avenida dos Astronautas 1758, S\~ao Jos\'e dos Campos, 12227-010, SP, Brazil}

\author{Jaziel G. Coelho}
\email{jazielcoelho@utfpr.edu.br}
\affiliation{Divis\~ao de Astrof\'isica, Instituto Nacional de Pesquisas Espaciais, Avenida dos Astronautas 1758, S\~ao Jos\'e dos Campos, 12227-010, SP, Brazil}
\affiliation{Departamento de F\'isica, Universidade Tecnol\'ogica Federal do Paran\'a, Medianeira, 85884-000, PR, Brazil}

\author{Jos\'e C. N. de Araujo}
\email{jcarlos.dearaujo@inpe.br}
\affiliation{Divis\~ao de Astrof\'isica, Instituto Nacional de Pesquisas Espaciais, Avenida dos Astronautas 1758, S\~ao Jos\'e dos Campos, 12227-010, SP, Brazil}

\begin{abstract}
Neutron stars (NSs) are excellent natural laboratories to constrain gravity on strong field regime and nuclear matter in extreme conditions.  Motivated by the recent discovery of a compact object with $2.59^{+0.08}_{-0.09} M_\odot$ in the binary merger GW190814, if this object was a NS, it serves as a strong constraint on the NS equation of state (EoS), ruling out several soft EoSs favored by GW170817 event. In this work, we revisit the question of the maximum mass of NSs considering a chameleon screening (thin-shell effect) on the NS mass-radius relation, where the microscopic physics inside the NS is given by realistic soft EoSs. We find that from appropriate and reasonable combination of modified gravity, rotation effects and realistic soft EoSs, that it is possible to achieve high masses and explain GW190814 secondary component, and in return also NSs like PSR J0740+6620 (the most NS massive confirmed to date). It is shown that gravity can play an important role in estimating maximum mass of NSs, and even with soft EoSs, it is possible to generate very high masses. Therefore, in this competition of hydrostatic equilibrium between gravity and pressure (from EoS choice), some soft EoSs, in principle, cannot be completely ruled out without first taking into account gravitational effects. 
\end{abstract}



\maketitle
\section{Introduction}

Recently, the LIGO/Virgo scientific collaboration announced the discovery of a compact binary coalescence, GW190814, involving a 22.2--24.3$M_\odot$ black hole (BH) and a compact object with a mass of 2.50--2.67$M_\odot$~\cite{2020ApJ...896L..44A}. The secondary component is either the most massive neutron star (NS), or the lightest BH ever discovered in a double compact-object system. The existence of ultra-massive NSs has been revealed in several studies. So far, the precise determination of the mass of pulsars leads to $M=1.97\pm0.04$~$M_\odot$ for PSR J1614--2230~\cite{2010Natur.467.1081D}, 
PSR J0348+0432 has $M=2.01\pm0.04$~$M\odot$~\cite{2013Sci...340..448A} and the highly likely the most massive NS observed to date, PSR J0740+6620, with $M=2.14\pm0.1$~$M_\odot$
~\cite{2020NatAs...4...72C}.

The discovery of the gravitational wave (GW) binary GW190814 triggered intense theoretical efforts about the real nature of its secondary component, in particular, the high mass requires the compact star matter to be described by a stiff equation of state (EoS). In \cite{Tsokaros} is argued that fast rotation is capable to explain the existence of a stable $\sim$2.6$M_\odot$ NS for moderately stiff EoS but may not be adequate for soft EoSs.  In particular, several soft EoSs favored by GW170817 and with maximum spherical masses of $\sim$2.1$M_\odot$ cannot be used to explain this source as a uniformly spinning NS. In \cite{Most} the authors infer a lower limit on the maximum mass $M_{\rm max}$ of non-rotating NSs, using arguments based on universal relations connecting the masses and spins of uniformly rotating NSs. Furthermore, they obtain a lower bound on the dimensionless spin for the secondary companion, using the upper maximum mass constraints from the GW170817 event \cite{2018ApJ...852L..25R,2019PhRvD.100b3015S}, and show that the allowed range in dimensionless spins correspond to rotational frequencies much higher than the fastest millisecond pulsars known \cite{2006Sci...311.1901H}. In \cite{2020arXiv200710999G} a rigorous upper bounds on maximum mass under the exclusive assumptions of causality and general relativity (GR), showing that the presence of a NS in GW190814 is not inconsistent with present observational constraints on the NS EoS. On the other hand, in \cite{2020arXiv200703799F} is showed that the stiffening of the EoS required to support ultra-massive NSs is inconsistent with either constraints obtained from the low deformability of medium-mass stars demanded by GW170817 or from energetic heavy-ion collisions. Several other methodology and speculations about GW190814 are presented in \cite{2020arXiv200708493D,2020arXiv200706526L,GW190814_01,GW190814_02,GW190814_03,GW190814_04,GW190814_05,GW190814_06,GW190814_07,GW190814_08,Lobato,GW190814_09,GW190814_10,GW190814_11,GW190814_12,GW190814_13,GW190814_14}.

As is well known, the two main observable of a NS, i.e., their mass and radius, both depend crucially on the choice of EoS and the gravitational theory, where gravity  will determine the macroscopic hydrostatic equilibrium. The general relativity is very well tested and compatible in weak-field observations, for instance, from solar system and terrestrial experiment tests \cite{Will,Williams,Adelberger}, but 
there are theoretical and observational reasons to believe that GR should be modified when gravitational fields are strong and/or on large scales. On large scales and from an observational point of view, the physical mechanism responsible for accelerating the Universe at late times is still an open question, and new degrees of freedom of the gravitational origin are alternatives to explain such an accelerated stage (see, e.g., \cite{DE_review,MG_review01} for review). Theories beyond GR can serve as alternatives to explain the current tension in the Hubble constant that persists in the framework of the standard cosmological model \cite{MG_H0_1,MG_H0_2,MG_H0_3}. Also, modified gravity models are motivated to drive the accelerating expansion of the Universe at early times, known as inflationary era. On astrophysical scales, 
compact objects such as BHs and NSs are our best natural laboratories to constrain strong gravity. In these bodies, gravity prevails over all other interactions and collapse leads to large-curvature and strong-gravity environments \cite{Psaltis}. We refer the reader to \cite{Berti} and references therein for several modified gravity scenarios motivation under the regime of strong gravitational field. 

A consistent way to modify gravity is to introduce some screening mechanism in order to have viable gravity, i.e., GR, at small distances and scales and possible relevant effects on others environmental scales, for instance, on compact objects or even on cosmological level. Recall that the screening mechanisms include chameleons \cite{Weltman1,Weltman2}, symmetrons \cite{Hinterbichler}, dilatons \cite{Brax}, Vainshtein mechanism \cite{Vainshtein}, etc. At the heart of screening mechanisms lies the fact that there are 29 orders-of-magnitude separating the cosmological and terrestrial densities and 20 orders of magnitude separating their distance scales. As a result, the properties of the new degrees of freedom of the gravitational origin (a scalar field) can vary wildly in different environments \cite{Burrage}. Screening gravity have been intensively investigated in the literature, in the most diverse aspects in cosmology and gravitation (see, e.g., \cite{SM01,SM02,SM03,SM04,SM05,SM06,SM07,SM08,SM09,SM10,SM11,SM12,SM13,SM14} for a short list). A compilation/list of various works on compact objects in modified theories of gravity can be found in \cite{Olmo}.  

As regards to NS observations, it seems there is a theoretical degeneration and is not clear if measurement of mass and radius  constrain possible gravity effects or EoSs. As showed in \cite{Eksi}, measurement of mass constrains gravity rather than the EoS [see also \cite{He}]. Thus, it is difficult to distinguish which of these effects is actually contributing to the actual observed mass and radius values. In this work, we analyze the NS mass-radius relation through a modification of the TOV equations induced by the presence of a possible chameleon screening (thin-shell effect), where the microscopic physics inside the NS will be modeled by realistic soft EoSs of which it is not possible to generate very massive NS with $\sim2.6$ $M_\odot$ or even $\sim2.14$ $M_\odot$ in GR context. Our main aim is to show that an appropriate combination of modified gravity, rotation effects and realistic soft EoSs, can have a joint effect for achieve high masses and explain GW190814 secondary component, and in return also NS like PSR J0740+6620.

This paper is organized as follows. In the next section, we present our theoretical framework and methodology to generate the NS mass-radius relation. In Section \ref{results} we present our main results and in Section \ref{final} our final comments and perspectives.
(The speed of light c is set equal to unity).

\section{Theoretical framework and Methodology}
\label{theory}

In this section we review in a nutshell the theoretical methodology used to obtain our main results.

\subsection{Screening Mechanisms}

The study of scalar-tensor theories has been motivated by some cosmological observation, especially in order to explain the current accelerating expansion stage of the Universe.  One proposed explanation is that gravity is modified on large scales, but must be suppressed on small scales, for instance, in the solar system deviations are constrained to be subdominant by a factor of $10^{-5}$. Screening gravity circumvent this problem by introducing non-linear modifications of the Poisson equation that dynamically suppress deviations from GR in the solar system without the need to fine-tune on the scalar mass or the coupling to matter. In short, screening mechanisms utilize non-linear dynamics to effectively decouple solar system and cosmological scales. At the heart of screening mechanisms lies the fact that there are 29 orders-of-magnitude separating the cosmological and terrestrial densities. As a result, the properties of the scalar field (new degree of freedom of gravitational origin) can vary widely in different environments.

In this work, we will consider the well-studied chameleon screening \cite{Weltman1,Weltman2}.
In chameleon models, the mass of the scalar is an increasing function of the ambient density. 
The purpose of this section is review the calculation of the parameterized post-Newtonian (PPN) parameter $\gamma$ that is relevant for chameleon theories. We follow the methodology presented in \cite{Sakstein} and references therein. 

Let us consider a specific subset of the general scalar-tensor theories, which in the Einstein frame is given by 
\begin{equation}
S=\int dx^4 \sqrt{-g}\left[\frac{M_{\rm pl}^2}{2}R-\frac{1}{2}\nabla_\mu\phi\nabla^\mu\phi -V(\phi)\right]+S_m[\tilde{g}],
\end{equation}
where the Jordan frame metric $\tilde{g}_{\mu \nu}$ is a Weyl rescaling of the Einstein frame metric $g_{\mu \nu }$ by a conformal factor $A(\phi)$, i.e., $\tilde{g}_{\mu \nu}=A^2(\phi)g_{\mu \nu}$. The coupling is given by

\begin{equation}
\alpha(\phi)=M_{\rm pl}\frac{d \ln A(\phi)}{d \phi}.
\end{equation}

Each particular scenario is set by the choice of $A(\phi)$ and $V(\phi)$ functions. The PPN metric for a single body, which we will refer to as body A with mass $M_A$, reads

\begin{align}\label{app:PPN_metric1}
\tilde{g}_{00}&=-1 +2 \frac{G^{PPN} M_A}{r}\\
\tilde{g}_{ij}&=\left(1+2\tilde{\gamma} \frac{G^{PPN} M_A}{r}\right)\delta_{ij},\label{app:PPN_metric2}
\end{align}

We will refer to the quantity $G^{PPN}$ as the PPN gravitational constant. $G^{PPN}$ controls the size of effects computed using the PPN metric. It is distinct from the gravitational constant $G$ that appears in GR and Newtonian gravity. We are interested in the regime where some body has some degree of screening. In this case, the equation inside of some screening radius is
\begin{equation}
\nabla^2\phi=0
\end{equation}
while outside the screening radius reads
\begin{equation}
\nabla^2\phi=8\pi \alpha(\phi) G\rho_A \,\,\,\,\, r > r_s,
\end{equation}
where $\alpha$ is the coupling function. 

In order to move on, we need to define $\alpha$. Here, we assume the chameleon field which uses a non-linear potential to make the field mass a function of the environmental density. The equation of motion reads
\begin{equation}
\nabla^2\phi=-n\frac{\Lambda^{4+n}}{\phi^{n+1}}+\frac{\alpha\rho_A}{M_{\rm pl}},
\end{equation}
where the effective potential is given by
\begin{equation}
V_{\rm eff}=\frac{\Lambda^{4+n}}{\phi^{n}}+\frac{\alpha \phi \rho_A}{M_{\rm pl}}.
\end{equation}

The mass-scale $\Lambda$ can vary over many orders of magnitude, but it is often compared to the dark energy scale, since this value is relevant for the present-day cosmic acceleration. Astrophysically, the chameleon profile of a spherically-symmetric object of mass $M$ and radius $R$ is not sourced by the object's mass, but rather by the mass inside a shell near the surface, a phenomenon that has been dubbed the thin-shell effect. The reason for this
is the following: deep inside the object, the field minimizes its effective potential corresponding to the ambient density but, as one moves away from the center, the field must eventually evolve asymptotically to the minimum at the density of the medium in which the object is immersed (astrophysical
or cosmological densities, depending on the situation to
be investigated). The field can only roll once the density is low enough so that its effective mass is light enough. The radius at which this happens is typically called the screening radius $r_{\rm s}$. 

The coupling function $\alpha$ above is a constant for chameleons \cite{Sakstein,Burrage}. Ignoring possible scalar mass contribution, the solution is then  

\begin{equation}
\phi=\phi_0-2\alpha^2\frac{Q_A G M_A}{r},
\end{equation}
where $Q$ is the scalar charge of body A and is given by
\begin{equation}
\label{eq:appcharge}
Q_A=\left(1-\frac{M_A(r_s^A)}{M_A}\right).
\end{equation}

Transforming to the Jordan frame and expanding $A(\phi)$ to first order in $GM_A/r$, one finds Eqs. \eqref{app:PPN_metric1} and \eqref{app:PPN_metric2} with
\begin{align}
\label{eq:GP} G^{PPN}&=G\left[1+2\alpha^2Q_A\right] \quad {\rm and} 
\\
\tilde{\gamma} & = \frac{1-2\alpha^2Q_A}{1+ 2\alpha^2Q_A}. \label{eq:gam1}
\end{align}

In a binary system, if we consider the orbital dynamics of a body of mass $M_B$ orbiting the body sourcing this metric, this second body may have its own screening radius $r_s^B$, so that the chameleon force between the two bodies $A$ and $B$ reads \cite{Mota}

\begin{equation}
F=-\frac{G M_A M_B}{r^2} \left[1+2 \alpha^2 Q_A Q_B \right]
\end{equation}
with the following modified potential
 
\begin{equation}\label{eq:Vcham}
V(r)=\frac{GM}{r}\left[1+2\alpha^2\left(1-\frac{M(r_{\rm s})}{M}\right)\right],
\end{equation}
where we have once again ignored the mass of the scalar. Thus, the physical quantity that is measurable in these theories is 
\begin{equation}
G^{PPN}=G \left[1+2 \alpha^2 Q_A Q_B \right],
\end{equation}
with corresponding PPN parameter $\gamma$ given by

\begin{equation}
\label{PPN_test}
\gamma =  2\left[1+2\alpha^2Q_AQ_B\right]^{-1}-1.
\end{equation}

Notice that other approaches have been developed in the literature, see for instance \cite{Burrage,SM10,SM11}. In what follows, let us quantify how this approximation presented above can modify spherical compact stars.

\subsection{Modified TOV}
In GR, the hydrostatic equilibrium of a star is described by the Tolman-Oppenheimer-Volkov (TOV) equations (see, e.g., \cite{ST1983}), namely

\begin{equation}
\label{dp_dr}
\frac{dP}{dr} = - \frac{G M(r) \rho(r)}{r^2}\Big(1 + \frac{P(r)}{\rho(r)} \Big) \Big(1 + \frac{4 \pi r^3 P(r)}{M(r)} \Big) \Big(1 - \frac{2GM(r)}{r} \Big)^{-1}
\end{equation}
and
\begin{equation}
\label{dm_dr}
\frac{dM}{dr} = 4 \pi r^2 \rho(r),
\end{equation}
where $\rho(r)$ is the energy density, $P(r)$ is the pressure and $M(r)$ is the mass within the radial coordinate $r$. 

The mass and the radius of a star are the two more obvious observables. Evidently, these quantities depends on the theory of gravitation and microscopic physics, i.e, the EoS of the matter of the star. In view of the formalism described in the previous section, we can notice that the chameleon screening can be quantified in the hydrostatic equilibrium rescaling $G$ by $G^{PPN}$ in Eqs. (\ref{dp_dr}) and (\ref{dm_dr}). $G^{PPN}$ is also directly  connected with the PPN parameter $\gamma$, since $\alpha^2 Q_A Q_B$ can be estimated. 

In this work, we will restrict to investigating NSs. Thus, we need to assume a EoS that describes the microscopic physics of the matter of  the NS to complete the system of equations above. Let us assume a realistic modeling of this physics, following the methodology described in \cite{eos}, where the nuclear matter inside NS are built by joining together different polytropic phases on a sequence of different density intervals (piecewise EoS), given by $\rho > \rho_0$, if for a set of dividing densities $\rho_0 < \rho_1 < \rho_2 ...$, the pressure and energy density are everywhere continuous and satisfy

\begin{equation}
\label{EOS}
P(\rho) = K_i \rho^{\Gamma_i}, \,\,\, d\Big(\frac{\epsilon}{\rho} \Big) =-P d\Big(\frac{1}{\rho}\Big), \,\,\,  \rho_{i-1} < \rho < \rho_i,
\end{equation}
where $\epsilon$ is the energy density and fixed by the first law of thermodynamics. 
The numerical modeling of EoSs is well summarized in \cite{eos} (see also \cite{eosii}). In this work, we will assume the low-density EoSs given by SLy, MS2, GNH3, ALF2, BBB2, H4, PS, WFF3, AP2 and FPS models. In total, we have 10 soft EoSs in our sample. In addition to several key properties of each ones, it should be stressed that these EoSs are consistent with constraints from GW170817 (see \cite{2017ApJ...850L..19M} and references therein). The best fit parametric configuration that describe these EoSs modeling are summarized in Table III in~\cite{eos}. 

On the other hand, if the NS is spinning, its maximum mass can be higher than $M_{\rm TOV}$ due to the additional rotational support against gravitational collapse to a BH. In particular, \cite{2016MNRAS.459..646B} computes the maximum mass allowed by uniform rotation, $M_{\rm max}$, purely in terms of the maximum mass of the non-rotating configuration. The importance of this universal relation, connecting the dimensionless spin on the stability line and the maximum dimensionless spin at the mass shedding limit, is that it allows us to calculate $M_{\rm max}$ sustainable through fast uniform rotation, finding that for any EoS it is about $20\%$ larger than the maximum mass supported by the corresponding non-rotating configuration (see \cite{Most,2016MNRAS.459..646B} for details). Thus, we complement our modeling considering $M_{\rm max} = \xi M_{\rm TOV}$, with $\xi = 1.203 \pm 0.022$ \cite{Most}.  Therefore, our full NS model will be constituted from some soft EoSs, rotation effects and chameleon screening. Note that this constraints on $\xi$ is obtained by assuming GR. In Section \ref{xi_in_MG}, we relax this condition and obtain $\xi$ within the theoretical framework here investigated. We will find $\xi=1.14^{+0.080}_{-0.082}$ in the modified gravity we investigated. Thus, these values are fully compatible with each other and this choice will not biased the main results to be presented below.

For the convenience of analysis, let us consider that there is a simple relationship between the expectation mass inside the screening radius in the form $M(r_{\rm s})_i = \beta_i M_i$, where $i$ runs over the bodies $A$ and $B$, in case of a binary system. In principle, we could consider more complicated dependency on the screening mass, for instance, like $M(r_{\rm s})_i = \beta_i r_i^n M_i$, where $r$ is the NS radius and $n$ some power law dependency. This type of correction will take PPN corrections predict more complicated relation/dependence on NS mass values. On the other hand, we expect that the screening radius be close to the surface of the star, so the approximation $M(r_{\rm s})_i = \beta_i M_i$ serves as a basis to quantify these screening effects within a simple approach, and to investigate possible new effects on the M-R relation. In what follows, we summarize our main results. 

\begin{table*}
\begin{center}
\setlength{\tabcolsep}{1em}
\renewcommand{\arraystretch}{2}
\begin{tabular}{c c c c c}
\hline
\hline
EOS & $M_{\rm max}$ (GR) & $M_{\rm max}$ (GR with rotation) & $M_{\rm max}$ (chameleon screening) & $M_{\rm max}$ (chameleon
screening with rotation)  \\
\hline
SLy  & 2.04 $M_{\odot}$ & 2.44 $M_{\odot}$ & 2.19$M_{\odot}$  &2.63$M_{\odot}$ \\
MS2  & 1.80 $M_{\odot}$ & 2.16 $M_{\odot}$ & 1.84$M_{\odot}$  &2.21$M_{\odot}$ \\
ALF4 & 1.94 $M_{\odot}$ & 2.32 $M_{\odot}$ & 2.06$M_{\odot}$  &2.48$M_{\odot}$ \\
GNH3 & 1.96 $M_{\odot}$ & 2.35 $M_{\odot}$ & 2.03$M_{\odot}$  &2.44$M_{\odot}$ \\
BBB2 & 1.91 $M_{\odot}$ & 2.29 $M_{\odot}$ & 2.07$M_{\odot}$  &2.49$M_{\odot}$ \\
H4  & 2.03 $M_{\odot}$ & 2.43 $M_{\odot}$ & 2.09$M_{\odot}$  &2.52$M_{\odot}$ \\
PS  & 1.75 $M_{\odot}$ & 2.10 $M_{\odot}$ & 1.80$M_{\odot}$  &2.10$M_{\odot}$ \\
WFF3 & 1.84 $M_{\odot}$ & 2.2 $M_{\odot}$ & 1.98$M_{\odot}$  &2.37$M_{\odot}$ \\
AP2 & 1.80 $M_{\odot}$ & 2.16 $M_{\odot}$ & 1.92$M_{\odot}$  &2.30$M_{\odot}$ \\
FPS & 1.79 $M_{\odot}$ & 2.14 $M_{\odot}$ & 1.93$M_{\odot}$  &2.32$M_{\odot}$ \\
\hline
\end{tabular}
\caption{Summary of the maximum mass values that can be obtained using 10 soft EoSs in different theoretical situations, namely, general relativity, general relativity with rapid uniform rotation effects, chameleon screening and chameleon screening with rapid uniform rotation effects.}
\label{tab:resultados_resumo}
\end{center}
\end{table*}

\begin{figure}
\begin{center}
\includegraphics[width=3.3in]{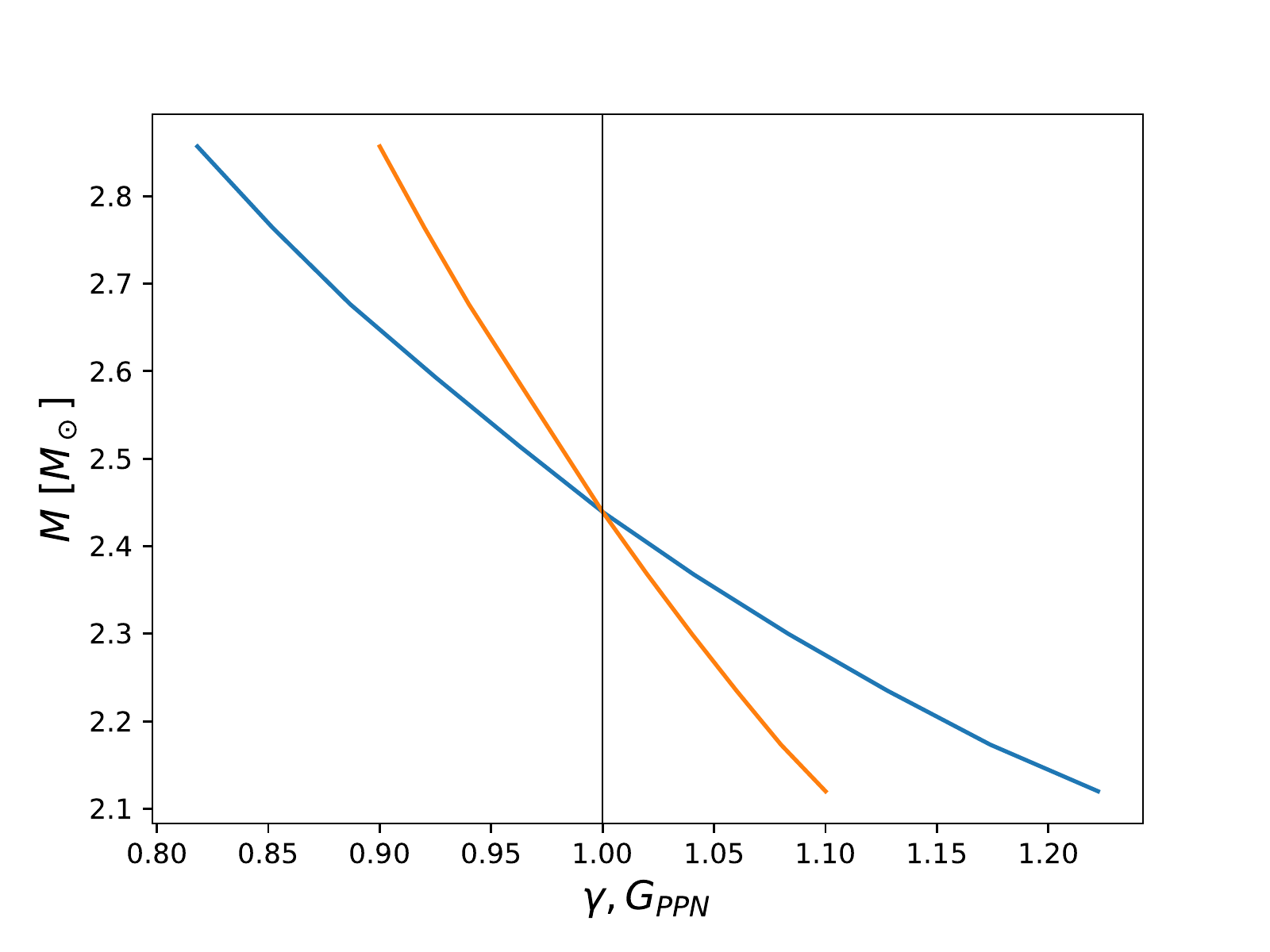}
\caption{Illustrative example of a neutron star maximum mass as a function of PPN parameter $\gamma$ (blue) and $G^{PPN}$ (orange) assuming EoS SLy.}
\label{M_PPN}
\end{center}
\end{figure}

\begin{figure*}
\begin{center}
\includegraphics[width=3.3in]{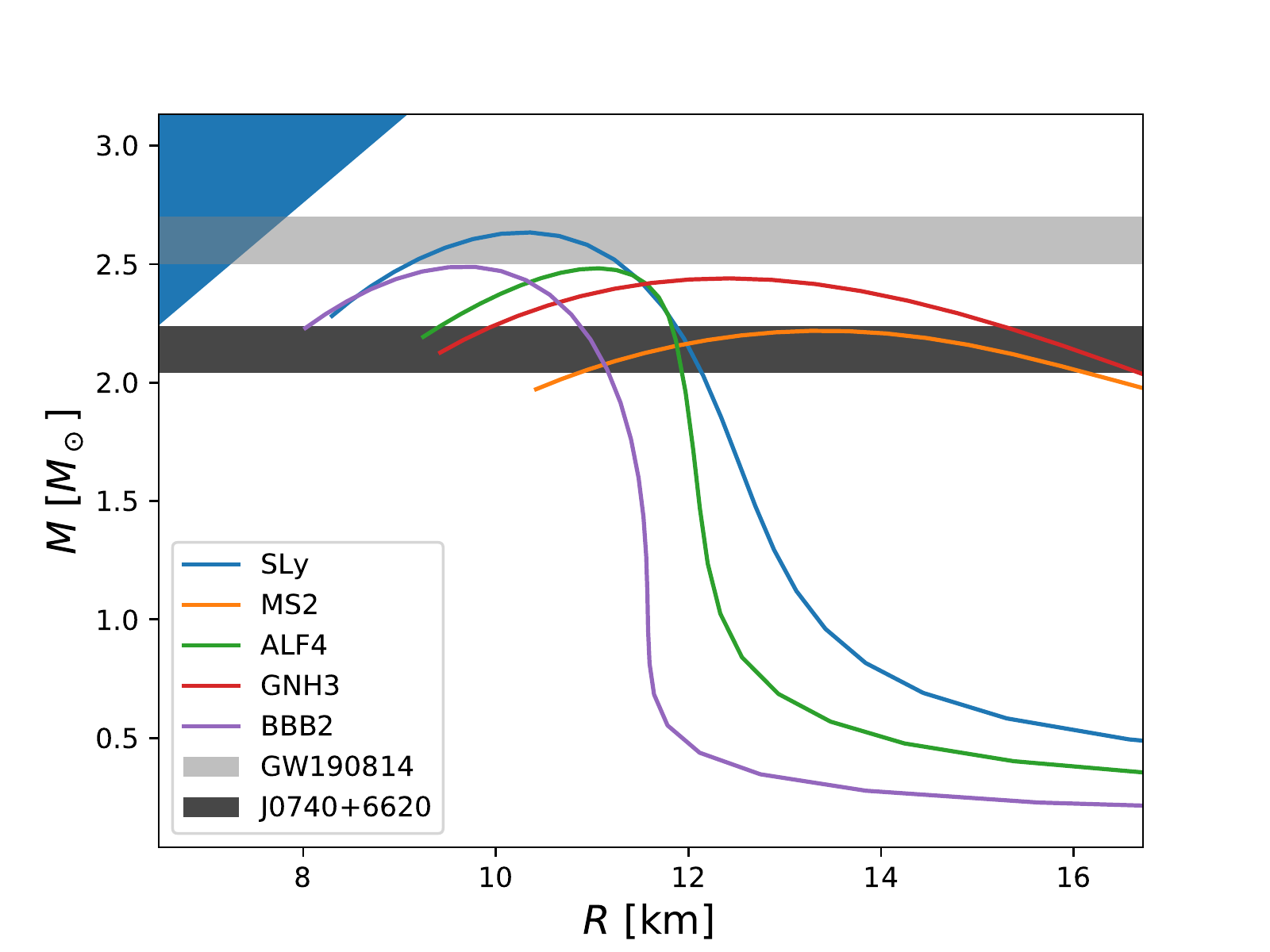} \,\,\,\,
\includegraphics[width=3.3in]{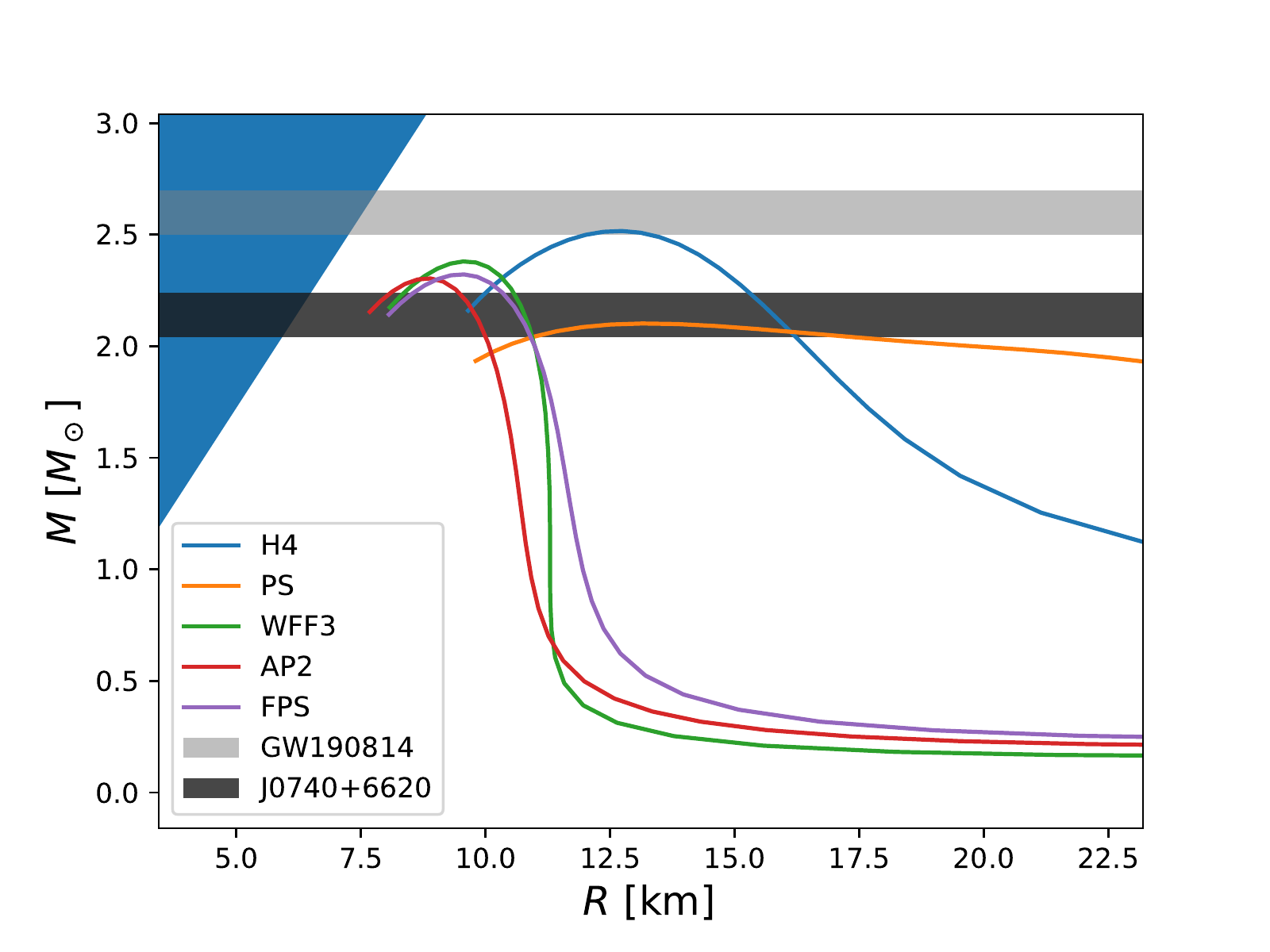}
\caption{Left panel: Neutron star mass-radius relation for different soft EoSs (SLy, MS2, ALF4, GNH3, BBB2) within Spinning + PPN corrections framework, with highlighted regions representing the neutron star mass expectation from the binary system  J0740+6620 and GW190814. Right panel: Same as in left panel, but from other soft EoSs (H4, PS, WFF3, AP2 and FPS).} 
\label{M_R_v}
\end{center}
\end{figure*}

\section{Results}
\label{results}

Table \ref{tab:resultados_resumo} summarizes the maximum mass of a NS taking into account 4 possible theoretical configurations and 10 input realistic soft EoSs, from which we can quantify how each ingredient is contributing to $M_{\rm max}$ estimates. Taking into account the screening mechanisms, we consider a NS with screening radius $r_s$ and $M_A$, orbiting a generic body B with $M_B$, so given the relation previously mentioned, $M(r_{\rm s})_i = \beta_i M_i$, we have $G^{PPN}/G = 1 + \bar{\alpha}$, with $\bar{\alpha} = 2\alpha^2(1 - \beta_B - \beta_A + \beta_B \beta_A)$, which is a global constant in our case. In all estimates in Table \ref{tab:resultados_resumo}, from the screening and screening + spinning cases, we assume a small and conservative total correction with $\bar{\alpha}=-0.05$. As we will see below (subsection A), through a statistical fit, effective corrections on $G^{PPN}$ are of this order of magnitude. i.e., $\bar{\alpha}$ $\sim$ -0.05 at $\sim$1$\sigma$. Corrections $\bar{\alpha} > 0$ tends to decrease the expectation value of maximum mass. Figure \ref{M_PPN} shows this using SLy EoS + spinning effects, quantifying how much $\gamma$ and $G^{\rm PPN}$ corrections can influence $M_{\rm max}$ prediction. Important to note that the expected corrections for $\gamma$ and $G^{\rm PPN}$ within screening gravity are not the same as measured on  earth or solar system. Without screening mechanisms, we would have to tune $\gamma$ value to satisfy terrestrial and solar system experiment bounds, but with screening mechanisms this bound can be automatically satisfied for this screened region with $M(r_s) \simeq  M(r)$ on these scales. Here, we are relaxing this condition in order to have new $\gamma$ and $G^{\rm PPN}$ expected values on a NS under thin-shell effect. Of course, any new $\gamma$ and $G^{\rm PPN}$ expected values should not deviate significantly from solar-system tests. In all our calculations, we are assuming a maximum deviation of 5\% when applied to NS structure equations.

Fig. \ref{M_R_v} shows the NS mass-radius relation from all the EOSs analyzed here, and summarized in Table \ref{tab:resultados_resumo} for spinning + chameleon screening combination, in direct comparison with the expected masses of the PSR J0740+6620, $2.14 \pm 0.10$ $M\odot$, a NS in a binary system with a white dwarf companion, and the mass of the lighter component of the GW190814 event, $2.6 \pm 0.10$ $M\odot$, if we interpret this object as being a NS. It is worth mentioning that in the results shown in Fig. \ref{M_R_v} and Table \ref{tab:resultados_resumo}, we assume an effective $\bar{\alpha}=-0.05$ (corresponding to an effective $G^{\rm PPN}/G = 0.95$).

The consideration above has an observational support. For instance, astrophysical tests on chameleon theories using distance indicators in the nearby universe are bounds to an effective values of $\Delta G/G \in [0.11 - 0.45]$ \cite{Jain}. The observation of GW170817 event imposed strong constraints on modified gravity/dark energy scenarios, inferring that the speed of GWs propagation is equal to the speed of light, so limiting the theoretical parametric space for various theories. Chameleon theories in principle has already this physical characteristic and this constraints does not affect in general any aspect in these theories. The authors in \cite{Wolf} calculate that the effects of a plausible variation in $G_N$ on the period decay of a local binary system are many orders of magnitude suppressed with respect to the effects of a change in $G_N$. As a result, $\dot{G}_N /G_N$ is constrained to $\sim 10^{-3}$, while $G_N$ is unconstrained by observations of the binary orbital decay rate. Here, the notation $G_N$ is the gravitational coupling between two matter sources. Note that our approach is linked to the  physical variable $G_N$ and not $\dot{G}_N$. In general, all our considerations about $G^{\rm PPN}/G$, i. e., all range of values assumed to the prior are compatible with the bounds derived from other observations.

\begin{figure}
\includegraphics[width=3.6in]{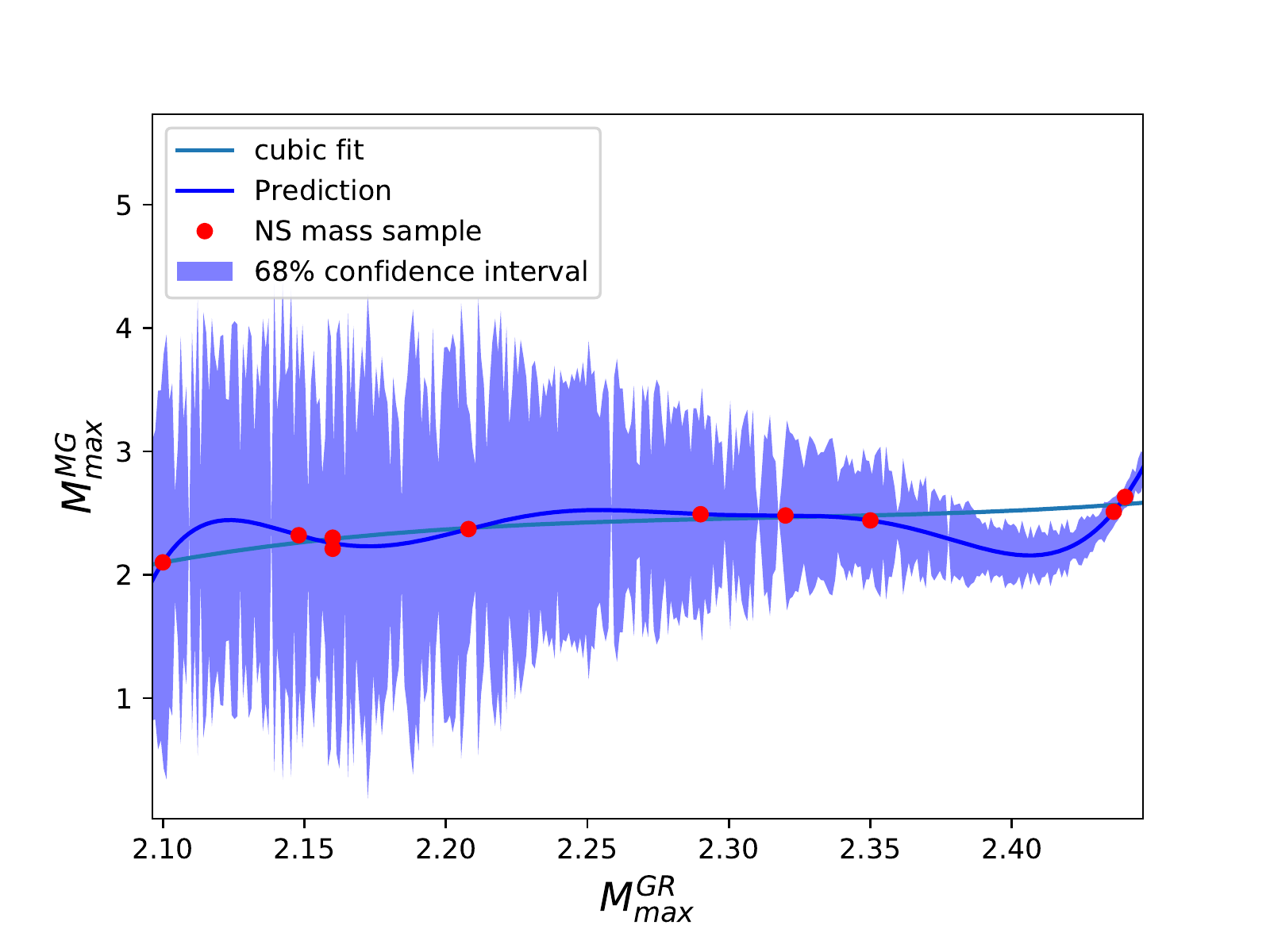} 
\caption{Statistical reconstruction in the plan $M^{MG}_{\rm max} - M^{GR}_{\rm max}$ using Gaussian processes for machine learning 
between the maximum mass estimates summarized in Table \ref{tab:resultados_resumo}. The solid blue curve is the Gaussian process mean. Both $M_{\rm max}$ are in units of $M_{\odot}$.}
\label{M_relation_ML}
\end{figure}

It is important to note that all EoSs assumed here share a relatively low maximum mass and thus they are called soft \cite{eos}. Soft EoSs consistent with GW170817 (see Table~\ref{tab:resultados_resumo}) are unable to provide enough mass to explain the secondary in GW190814, even for a NS endowed with maximum uniform rotation \cite{Tsokaros}. We noticed that assuming non-rotating mass configuration, it is troublesome to predicting high mass in GR, and all these EoSs have difficulties in reaching the mass expected by J0740+6620. 

Thus, in this simplest case, all of these EoSs are excluded. But, adding rotation effects, all these EoSs can explain J0740+6620 easily. The authors in \cite{1994ApJ...424..823C,1994ApJ...422..227C} were the first to show that spinning up a NS uniformly can increase its mass by up to $\sim20\%$. Thus, soft EoSs consistent with rapid uniform rotation is enough to explain that compact object. The analysis frame change the situation if we look at GW190814 secondary component. When considering GR + spinning, all these soft EOSs can not achieve very high masses, and all them should be ruled out. This means that the secondary compact object in GW190814 can not be explained by an EoS like SLy (EoS favored by GW170817). Also, as argued in \cite{Tsokaros}, such EoSs like SLy must now be rejected because their mass-shedding limit is below the lower limit mass of this object in GW190814. But, we find that modifying the hydrostatic equilibrium equations of NSs, incorporating a possible thin-shell effect (in our study from chameleon screening), we can reach larger masses. We note that chameleon screening can achieve high masses and explain GW190814 with soft EoSs like SLy, BB2, ALF4 and H4 (see Fig. \ref{M_R_v}), which in GR context is not possible. The BBB2, ALF4 and H4 EoSs predictions live in the lower limit of the error bar of GW190814, while the SLy EoS easily archive higher values and can explain GW190814 best fit and more massive objects.
Therefore, in presence of new degrees of freedom of gravitational origin like a chameleon screening, it is possible to maintain these soft EoSs to describe the internal structure of these compact objects. 

Note that the same EoS in screening + spinning context can explain both NSs in J0740+6620 and GW190814 at the same time, changing only the radius prediction. For instance, from EoS SLy analysis, one obtains [9.16 - 11.27]km and [10.86 - 11.18]km for GW190814 and J0740+6620, respectively. 
On the other hand, it is interesting to derive a general relation between the maximum NS mass from GR and the modified gravity theory from some EoS sample. We will obtain this relationship based on our 10 soft EoSs sample \cite{eos}, where the main results are summarized in Table \ref{tab:resultados_resumo}.

To get this relation, let us use a Gaussian Processes (GP).  The GP consists of generic supervised learning method designed to solve regression and probabilistic classification problems, where we can
interpolate the observations and compute empirical confidence intervals and a prediction in some region of interest \cite{ML}. Let us consider that our 10 estimates in Table \ref{tab:resultados_resumo} (the case with rotation effects) are data points to be trained between a certain mass range. Note that 10 data points, from 10 different soft EoSs, give us reasonable information to get a good prediction within the mass range where soft EoS there due the robustness of this methodology. The GP method is the state-of-the-art to obtain statistical information and model prediction from some previously known information or data.

Figure \ref{M_relation_ML} show the data reconstruction. The solid blue curve is the GP mean prediction and the shaded areas is the statistical confidence level (CL) at 68\%. Each red point represents the mass information (data point) in the plane $M^{MG}_{\rm max} - M^{GR}_{\rm max}$, in the case with rotation, which has been trained to obtain the reconstruction and prediction between these mass estimates. During the reconstruction, we use the most popular covariance functions in the literature, the standard Gaussian Squared-Exponential. To obtain these results, we make use of some numerical routines available in \cite{scikit-learn}. Interesting to note that from a soft EoS sample, in general, we can obtain high mass, like a 3 solar mass, within modified gravity approach even at 68\% CL.

In Figure \ref{M_relation_ML}, we also show a cubic fit that best fit the plan $M^{MG}_{\rm max} - M^{GR}_{\rm max}$ in the mass range $M \in [2.10, 2.46]$ $M_{\odot}$ in GR prediction. We find

\begin{equation}
M^{MG}_{\rm max}= -320.38 + 417.26 M_{\rm max} - 179.96 M^{2}_{\rm max} + 25.90 M^{3}_{\rm max},
\end{equation}
where $M_{\rm max}$ is the maximum mass in GR.

Thus, within the theoretical framework investigated here, Figure \ref{M_relation_ML}, summary the most general relations and prediction analysis between $M^{MG}_{\rm max}$ and $M^{GR}_{\rm max}$ within the soft EoS approach.

\begin{figure}
\begin{center}
\includegraphics[width=3.1in]{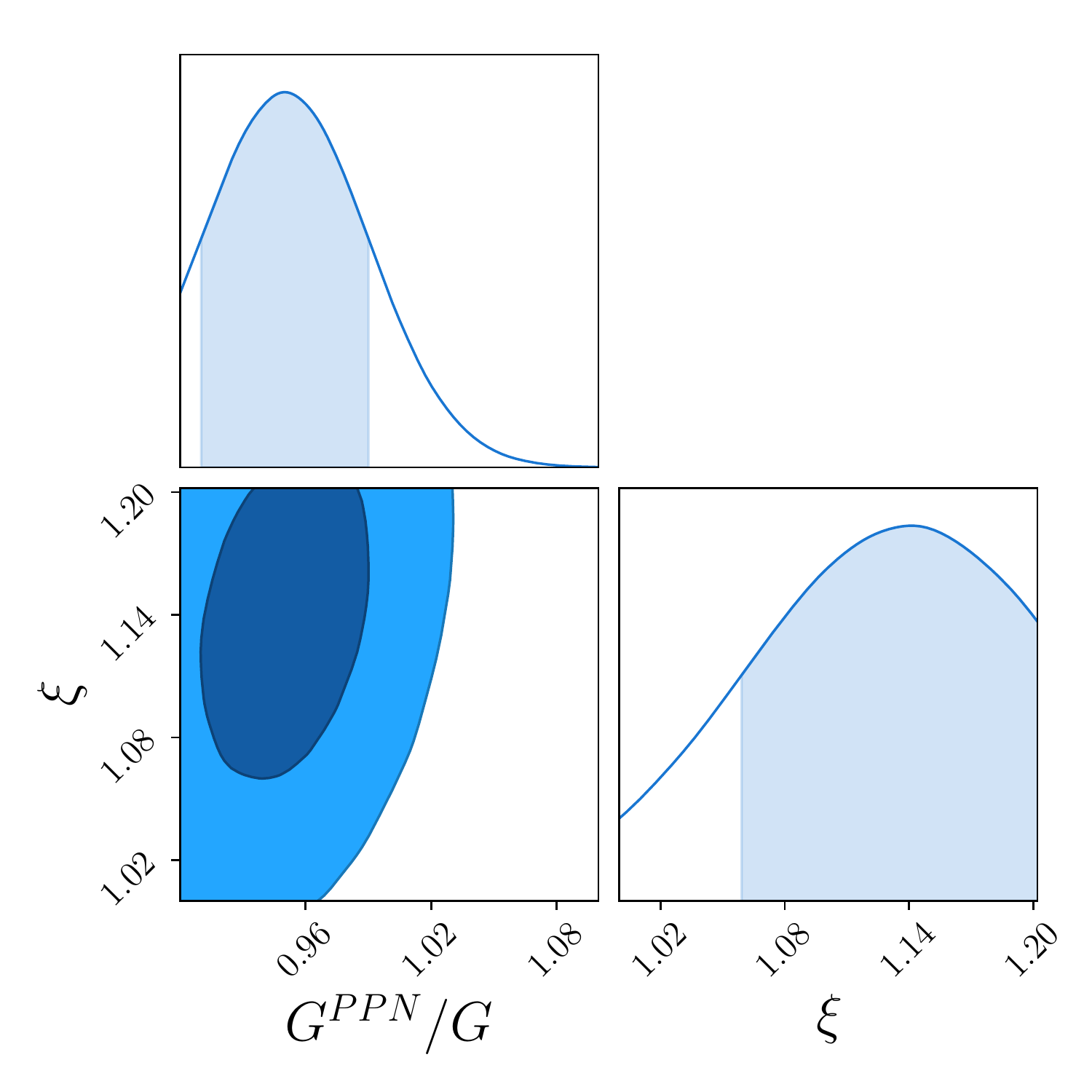} \,\,\,\,
\caption{Two-dimensional marginalized distributions in the parametric space in the plan $G^{\rm PPN}/G - \xi$ at 68\% CL and 95\% CL.}
\label{PS_xi_G}
\end{center}
\end{figure}

\subsection{Rotation effects in modified gravity}
\label{xi_in_MG}

The spin of the secondary object is a quantity that has not been constrained by the observations of GW190814. Furthermore, we know that the condition for rapid uniform rotation contributes significantly to increase the predicted maximum mass. 
In the results previously presented, we assume a rotation value that was obtained in GR. Once that the gravity theory has been modified, it is necessary to re-calculate the rotational effect in the new gravity framework. Let us obtain a new estimate on $\xi$, in the situation we can still explain GW190814 secondary component, if this compact object was a NS, within the screening + rotation framework. For this, we carry out a simple $\chi^2$ fit given by

\begin{equation}
\chi^2 = \frac{(M_O - M_{\rm max}(G^{\rm PPN}, \xi))^2}{\sigma^2_O},
\end{equation}
where $M_{\rm max}(G^{\rm PPN}, \xi)$ is the maximum mass expected assuming $G^{\rm PPN}$, $\xi$ as free parameter, where we assume the prior $\xi \in$ [1, 1.2]. The range in $\xi$ represent non-rotating with $\xi = 1$ and maximum rotation with $\xi = 1.2$. This upper prior on $\xi$ can be interpreted that 20\% is probably no longer applicable and this correction is an upper limit.
The quantities $M_O$, $\sigma^2_O$, are the mass and the associate error bar at 1$\sigma$ confidence level (CL) from the GW190814 secondary component observation. 

Note that in this approach we are obtaining a new $\xi$ constraints parametrically through the relation $M^{MG}_{\rm max} = \xi M^{MG}_{\rm TOV}$. Figure \ref{PS_xi_G} shows the parametric space in the plan $G^{\rm PPN}/G - \xi$ at 68\% CL and 95\% CL, after assuming $\xi$ as a free parameter. We find $G^{\rm PPN}/G=0.951^{+0.039}_{-0.041}$ (with the corresponding $\gamma =0.90 \pm 0.093$) and $\xi=1.14^{+0.080}_{-0.082}$. During the analysis we assume SLy as input EoS. Note that this value is statistically compatible with the value inferred in GR. Thus, we conclude that the values and analysis derived in the previous section are not biased from this perspective.

On the other hand, other independent measurements, from other observational perspectives, have been carried out recently reporting, $\gamma=0.87^{+0.19}_{-0.17}$ \cite{gamma1} and $|\gamma -1| < 0.2 (\Lambda/100)$kpc \cite{gamma2}, with $\Lambda =10 -200$kpc. Our constraints are in total agreement with these values. All these constraints suggests no deviation from GR. But, despite the statistical bounds be compatible with GR, note that the gravitational relaxing induced by $G^{\rm PPN}$ is enough to generate NS with $M \sim 2.6$ $M_\odot$ or more, using a soft EoS. In this case, the NS can rotate below of the maximum limit, which in turn will generate an effective $G^{\rm PPN}$ slightly smaller, but statistically equivalent even at 68\% CL when considering the maximum value. From Eq.(\ref{eq:GP}), if we assume the coupling parameter to be $\alpha \simeq 1$ \cite{Burrage,Sakstein}, we find $M(r_s)/M \simeq 0.98$, showing that the screening mass is near to the surface, as expected.
\\

\section{Final Remarks}
\label{final}

We consider that the gravitational interaction can deviate minimally from that predicted from GR, through a PPN correction induced
by a thin-shell effect via a chameleon screening. We analyze its impacts on the NS mass-radius relation, motivated to check if it is possible to generate high NS masses (when compared to GR prediction). Our main conclusion is that from a combination of modified gravity, rotation effects and realistic soft EoSs, it is possible to achieve high masses like $M \sim 2.6$ $M_\odot$ or more, and explain GW190814 secondary component. In return, it is also possible to explain
the most NS massive confirmed to date, i.e., PSR J0740+6620. 
Then one sees here the following interesting aspect: one can not ruled out some soft EoSs without first taking into account effects coming from alternative theories of gravity. Although our theoretical approach be simple, this consequence is clear (see, e.g., Table \ref{tab:resultados_resumo} and Figure \ref{M_relation_ML}).  Therefore, gravity can play an important role, without the need to invoke very stiff EoSs or unusual aspects in nuclear matter.  

It might be interesting to see how more sophisticated gravity theories behave in this perspective, and if it is possible to obtain new and accurate observational bounds on the free parameters that characterize such scenarios using possible massive NS observations. 

\begin{acknowledgments}
\noindent 
The authors thank the referee for comments which helped to improve the quality of the manuscript. RCN would like to thank the agency FAPESP for financial support under the projects 2018/18036-5 and 2013/26258-4. JGC is likewise grateful to the support of CNPq (421265/2018-3 and 305369/2018-0). JCNA thanks the partial support of CNPq (308367/2019-7).
\end{acknowledgments}


\begin{thebibliography}{}

\bibitem{2020ApJ...896L..44A} R. Abbottet et al, \textit{GW190814: Gravitational Waves from the Coalescence of a 23 Solar Mass Black Hole with a 2.6 Solar Mass Compact Object}. The Astrophysical Journal {\bf 896}, L44 (2020)
\href{https://arxiv.org/abs/2006.12611}{2006.12611}

\bibitem{2010Natur.467.1081D} P. Demorest, T. Pennucci, S. Ransom, M. Roberts, J. Hessels, 
\textit{A two-solar-mass neutron star measured using Shapiro delay}.
Nature {\bf 467}, 1081–1083 (2010)
\href{https://arxiv.org/abs/1010.5788}{1010.5788}

\bibitem{2013Sci...340..448A} J. Antoniadis et al \textit{A Massive Pulsar in a Compact Relativistic Binary}.\ Science {\bf 340}, 448 (2013)
\href{https://arxiv.org/abs/1304.6875}{1304.6875}

\bibitem{2020NatAs...4...72C} H. Cromartie et al, \textit{Relativistic Shapiro delay measurements of an extremely massive millisecond pulsar}.\ Nature Astronomy {\bf 4}, 72–76 (2020)
\href{https://arxiv.org/abs/1904.06759}{1904.06759}

\bibitem{Tsokaros} A. Tsokaros, M. Ruiz and S. L. Shapiro,
\textit{GW190814: Spin and equation of state of a neutron star companion}, (2020). \href{https://arxiv.org/abs/2007.05526}{2007.05526} 

\bibitem{Most} E. R. Most, L. J. Papenfort, L. R. Weih and L. Rezzolla,
\textit{A lower bound on the maximum mass if the secondary in GW190814 was once a rapidly spinning neutron star}, (2020) \href{https://arxiv.org/abs/2006.14601}{2006.14601}

\bibitem{2019PhRvD.100b3015S} M. Shibata, E. Zhou, K. Kiuchi, S. Fujibayashi, \textit{Constraint on the maximum mass of neutron stars using GW170817 event}.\ Physical Review D {\bf 100}, 023015 (2019)
\href{https://arxiv.org/abs/1905.03656}{1905.03656}

\bibitem{2018ApJ...852L..25R} L. Rezzolla, E. Most, and L. Weih, \textit{Using Gravitational-wave Observations and Quasi-universal Relations to Constrain the Maximum Mass of Neutron Stars}.\ The Astrophysical Journal {\bf 852}, L25 (2018)
\href{https://arxiv.org/abs/1711.00314}{1711.00314}


\bibitem{2006Sci...311.1901H} J. Hessels et al, \textit{A Radio Pulsar Spinning at 716 Hz}.\ Science {\bf 311}, 1901–1904 (2006)
\href{https://arxiv.org/abs/astro-ph/0601337}{0601337}

\bibitem{2020arXiv200710999G} A. Godzieba, D. Radice, S. Bernuzzi \textit{On the maximum mass of neutron stars and GW190814}.
\href{https://arxiv.org/abs/2007.10999}{2007.10999}

\bibitem{2020arXiv200703799F} F. Fattoyev, J. Horowitz, J. Piekarewicz, and J. Reed,
\textit{GW190814: Impact of a 2.6 solar mass neutron star on nucleonic equations of state},
\href{https://arxiv.org/abs/2007.03799}{2007.03799}

\bibitem{2020arXiv200708493D} V. Dexheimer, R. Gomes, T. Kl{\"a}hn, S. Salinas 
\textit{GW190814 as a massive rapidly-rotating neutron star with exotic degrees of freedom}.
\href{https://arxiv.org/abs/2007.08493}{2007.08493}

\bibitem{2020arXiv200706526L} Y. Lim, A. Bhattacharya, J. Holt, D. Pati,
\textit{Revisiting constraints on the maximum neutron star mass in light of GW190814}
\href{https://arxiv.org/abs/2007.06526}{2007.06526}

\bibitem{GW190814_01} S.Clesse and J. G. Bellido,
\textit{GW190425 and GW190814: Two candidate mergers of primordial black holes from the QCD epoch}, \href{https://arxiv.org/abs/2007.06481}{2007.06481}

\bibitem{GW190814_02} I. Tews et al, 
\textit{On the nature of GW190814 and its impact on the understanding of supranuclear matter}, \href{https://arxiv.org/abs/2007.06057}{2007.06057}

\bibitem{GW190814_03}  H. Tan, J. N. Hostler, and N. Yunes
\textit{Kinky neutron stars in light of GW190814}, \href{https://arxiv.org/abs/2006.16296}{2006.16296}

\bibitem{GW190814_04} N.B. Zhang and B. A. Li,
\textit{GW190814's secondary component with mass (2.50-2.67) as a super-fast pulsar}, \href{https://arxiv.org/abs/2007.02513}{2007.02513}

\bibitem{GW190814_05} Z. Roupas, G. Panotopoulos and I. Lopes,
\textit{QCD color superconductivity in compact stars: color-flavor locked quark star candidate for the gravitational-wave signal GW190814}, \href{https://arxiv.org/abs/2010.11020}{2010.11020}

\bibitem{GW190814_06} A. V. Astashenok, S. Capozziello, S. D. Odintsov and V. K. Oikonomou,
\textit{Extended Gravity Description for the GW190814 Supermassive Neutron Star}, \href{https://arxiv.org/abs/2008.10884}{2008.10884}

\bibitem{GW190814_07} A.~V.~Astashenok and S.~D.~Odintsov,
A.~V.~Astashenok, S.~Capozziello and S.~D.~Odintsov,
\textit{Extreme neutron stars from Extended Theories of Gravity}, JCAP {\bf 1501} (2015) 001, arXiv:1408.3856.

\bibitem{GW190814_08} C. E. Mota, L. C. N. Santos, F. M. da Silva, G. Grams, I. P. Lobo and D. P. Menezes,
\textit{Generalized Rastall's gravity and its effects on compact objects}, \href{https://arxiv.org/abs/2007.01968}{2007.01968}

\bibitem{Lobato} R. Lobato et al, \textit{Neutron stars in f(R,T) gravity using realistic equations of state in the light of massive pulsars and GW170817}, \href{https://arxiv.org/abs/2009.04696}{2009.04696}

\bibitem{GW190814_09} V. I. Danchev and D. D. Doneva,
\textit{Constraining the equation of state in modified gravity via universal relations}, \href{https://arxiv.org/abs/2010.07392}{2010.07392}

\bibitem{GW190814_10} Z. Roupas, 
\textit{Secondary component of gravitational-wave signal GW190814 as an anisotropic neutron star}, \href{https://arxiv.org/abs/2007.10679}{2007.10679}

\bibitem{GW190814_11} I. Bombaci, A. Drago, D. Logoteta, G. Pagliara and I. Vidana,
\textit{Was GW190814 a black hole -- strange quark star system?}, \href{https://arxiv.org/abs/2010.01509}{2010.01509}

\bibitem{GW190814_12} V. De Luca, V. Desjacques, G. Franciolini, P. Pani, A. Riotto,
\textit{GW190814: Circumstantial Evidence for Up-Down Quark Star}, \href{https://arxiv.org/abs/2009.00942}{2009.00942}

\bibitem{GW190814_13} Z. Cao, L. W. Chen, P. C. Chu and Y. Zhou,
\textit{The GW190521 Mass Gap Event and the Primordial Black Hole Scenario}, \href{https://arxiv.org/abs/2009.01728}{2009.01728}

\bibitem{GW190814_14} K. Vattis, I. S. Goldstein and S. M. Koushiappas,
\textit{Could the 2.6$M_\odot$ object in GW190814 be a primordial black hole?}, \href{https://arxiv.org/abs/2006.15675}{2006.15675}

\bibitem{Will} C. Will, 
\textit{The Confrontation between General Relativity and Experiment}, \href{https://link.springer.com/article/10.12942\%2Flrr-2014-4}{Living Rev. Relativity {\bf 17} 4 (2014)}. \href{https://arxiv.org/abs/1403.7377}{1403.7377}

\bibitem{Williams} J. G. Williams, S. G. Turyshev, D. H. Boggs, 
\textit{Progress in lunar laser ranging tests of relativistic gravity}, \href{https://doi.org/10.1103/PhysRevLett.93.261101}{Phys Rev Lett {\bf 93} 261101, (2004) }. \href{https://arxiv.org/abs/gr-qc/0411113}{0411113}

\bibitem{Adelberger} E. G. Adelberger, B. Heckel, S. Hoedl, C. Hoyle, D. Kapner, and A. Upadhye, 
\textit{Particle physics implications of a recent test of the gravitational inverse sqaure law}, \href{https://doi.org/10.1103/PhysRevLett.98.131104.}{Phys Rev Lett {\bf 98} 131104, (2007)}. \href{https://arxiv.org/abs/hep-ph/0611223}{0611223}

\bibitem{DE_review} D. Huterer and D. L. Shafer, Rep. Prog. Phys. {\bf 81}, 016901 (2018), arXiv:1709.01091.

\bibitem{MG_review01} M. Ishak,  Living Rev. Rel. {\bf 1}, 22 (2019), arXiv:1806.10122.

\bibitem{MG_H0_1} M. Ballardini, M. Braglia, F. Finelli, D. Paoletti and A. A. Starobinsky, arXiv:2004.14349.
\textit{Scalar-tensor theories of gravity, neutrino physics, and the H0 tension},  \href{https://arxiv.org/abs/2004.14349}{2004.14349}

\bibitem{MG_H0_2} H. Desmond, B. Jain and J. Sakstein,
\textit{A local resolution of the Hubble tension: The impact of screened fifth forces on the cosmic distance ladder}, \href{https://journals.aps.org/prd/abstract/10.1103/PhysRevD.100.043537}{Phys. Rev. D {\bf 100}, 043537 (2019)}. \href{https://arxiv.org/abs/1907.03778}{1907.03778}

\bibitem{MG_H0_3} R. D'Agostino and R. C. Nunes, 
\textit{Measurements of H0 in modified gravity theories: The role of lensed quasars in the late-time Universe}, \href{https://journals.aps.org/prd/abstract/10.1103/PhysRevD.101.103505}{Phys. Rev. D {\bf 101}, 103505 (2020)}. \href{https://arxiv.org/abs/2002.06381}{2002.06381}


\bibitem{Psaltis} D. Psaltis,
\textit{Probes and Tests of Strong-Field Gravity with Observations in the Electromagnetic Spectrum}, \href{https://link.springer.com/article/10.12942\%2Flrr-2008-9}{Living Rev. Relativ. {\bf 11} (2008)}. \href{https://arxiv.org/abs/0806.1531}{0806.1531}

\bibitem{Berti} E. Berti {\it et al.,}, 
\textit{Testing General Relativity with Present and Future Astrophysical Observations}, \href{https://iopscience.iop.org/article/10.1088/0264-9381/32/24/243001}{Class. Quantum Grav. {\bf 32}, 243001 (2015)}. \href{https://arxiv.org/abs/1501.07274}{1501.07274}

\bibitem{Weltman1} J. Khoury and A. Weltman,
\textit{Chameleon Fields: Awaiting Surprises for Tests of Gravity in Space}, \href{https://journals.aps.org/prl/abstract/10.1103/PhysRevLett.93.171104}{Phys. Rev. Lett. \textbf{93} 171104, (2004)}. \href{https://arxiv.org/abs/astro-ph/0309300}{0309300}

\bibitem{Weltman2} J. Khoury and A. Weltman,
\textit{Chameleon Cosmology}, \href{https://journals.aps.org/prd/abstract/10.1103/PhysRevD.69.044026}{Phys. Rev. D \textbf{69} 044026, (2004)}. \href{https://arxiv.org/abs/astro-ph/0309411}{0309411}

\bibitem{Hinterbichler} K. Hinterbichler and J. Khoury, 
\textit{Symmetron Fields: Screening Long-Range Forces Through Local Symmetry Restoration}, \href{https://journals.aps.org/prl/abstract/10.1103/PhysRevLett.104.231301}{Phys. Rev. Lett. {\bf 104},
231301 (2010)}. \href{https://arxiv.org/abs/1001.4525}{1001.4525}

\bibitem{Brax} P. Brax, C. van de Bruck, A. C. Davis, and D. Shaw, 
\textit{The dilaton and modified gravity.}, \href{https://doi.org/10.1103/PhysRevD.82.063519}{Phys Rev D {\bf 82} 063519 (2010)}. \href{https://arxiv.org/abs/1005.3735}{1005.3735}

\bibitem{Vainshtein} A. I. Vainshtein, Phys. Lett. 39B, 393 (1972).

\bibitem{Burrage} C. Burrage and J. Sakstein, 
\textit{Tests of Chameleon Gravity}, \href{https://link.springer.com/article/10.1007\%2Fs41114-018-0011-x}{Living Rev. Relativity \textbf{21} 1 (2018)}. \href{https://arxiv.org/abs/1709.09071}{1709.09071}

\bibitem{SM01} P. Brax, A. C. Davis, and R. Jha,
\textit{Neutron Stars in Screened Modified Gravity: Chameleon vs Dilaton
}, \href{https://journals.aps.org/prd/abstract/10.1103/PhysRevD.95.083514}{Phys. Rev. D {\bf 95}, 083514 (2017)}. \href{https://arxiv.org/abs/1702.02983}{1702.02983}

\bibitem{SM02} A. Dima and F. Vernizzi,
\textit{Vainshtein Screening in Scalar-Tensor Theories before and after GW170817: Constraints on Theories beyond Horndeski
}, \href{https://journals.aps.org/prd/abstract/10.1103/PhysRevD.97.101302}{Phys. Rev. D {\bf 97}, 101302 (2018)}. \href{https://arxiv.org/abs/1712.04731}{1712.04731}

\bibitem{SM03} D. F. Mota and D. J. Shaw,
\textit{Evading Equivalence Principle Violations, Cosmological and other Experimental Constraints in Scalar Field Theories with a Strong Coupling to Matter
}, \href{https://journals.aps.org/prd/abstract/10.1103/PhysRevD.75.063501}{Phys. Rev. D {\bf 75} 063501, (2007)}. \href{https://arxiv.org/abs/hep-ph/0608078}{0608078}

\bibitem{SM04} X. Zhang, T. Liu, and W. Zhao,
\textit{Gravitational radiation from compact binary systems in screened modified gravity
}, \href{https://journals.aps.org/prd/abstract/10.1103/PhysRevD.95.083514}{Phys. Rev. D {\bf 95}, 104027 (2017)}. \href{https://journals.aps.org/prd/abstract/10.1103/PhysRevD.95.104027}{1702.08752}

\bibitem{SM05} P. Brax, A. C.Davis, B. Li, and H. A. Winther,
\textit{A Unified Description of Screened Modified Gravity
}, \href{https://journals.aps.org/prd/abstract/10.1103/PhysRevD.95.083514}{Phys. Rev. D {\bf 95} 083514, (2017)}. \href{https://arxiv.org/abs/1203.4812}{1203.4812}

\bibitem{SM06} H. Desmond and J. Sakstein,
\textit{Screened fifth forces lower the TRGB-calibrated Hubble constant too
}, \href{https://journals.aps.org/prd/abstract/10.1103/PhysRevD.86.044015}{Phys. Rev. D {\bf 86} 044015, (2012)}. \href{https://arxiv.org/abs/2003.12876}{2003.12876}

\bibitem{SM07} P. Hamilton, M. Jaffe, P. Haslinger, Q. Simmons, H. Müller, J. Khoury,
\textit{Atom-interferometry constraints on dark energy
}, \href{https://science.sciencemag.org/content/349/6250/849}{Science, {\bf 349} (2015)}. \href{https://arxiv.org/abs/1502.03888}{1502.03888}

\bibitem{SM08} D. O. Sabulsky, I. Dutta, E. A. Hinds, B. Elder, C. Burrage, E. J. Copeland,
\textit{Experiment to detect dark energy forces using atom interferometry
}, \href{https://journals.aps.org/prl/abstract/10.1103/PhysRevLett.123.061102}{Phys. Rev. Lett. {\bf 123}, 061102 (2019)}. \href{https://arxiv.org/abs/1812.08244}{1812.08244}

\bibitem{SM09} C. Renevey, J. Kennedy, and L. Lombriser
\textit{Parameterised post-Newtonian formalism for the effective field theory of dark energy via screened reconstructed Horndeski theories
}, \href{https://arxiv.org/abs/2006.09910}{2006.09910}

\bibitem{SM10} X. Zhang, W. Zhao, H. Huang, and Y. Cai,
\textit{Post-Newtonian parameters and cosmological constant of screened modified gravity
}, \href{https://journals.aps.org/prd/abstract/10.1103/PhysRevD.93.124003}{Phys. Rev. D {\bf 93}, 124003 (2016)}. \href{https://arxiv.org/abs/1603.09450}{1603.09450}

\bibitem{SM11} A. Hees, A. Füzfa,
\textit{Combined cosmological and solar system constraints on chameleon mechanism
}, \href{https://journals.aps.org/prd/abstract/10.1103/PhysRevD.85.103005}{Phys. Rev. D {\bf 85}, 103005, 2012}. \href{https://arxiv.org/abs/1111.4784}{1111.4784}

\bibitem{SM12} K. Koyama and J. Sakstein,
\textit{Astrophysical Probes of the Vainshtein Mechanism: Stars and Galaxies}, \href{https://journals.aps.org/prd/abstract/10.1103/PhysRevD.91.124066}{Phys. Rev. D {\bf 91}, 124066 (2015)}. \href{https://arxiv.org/abs/1502.06872}{1502.06872}

\bibitem{SM13} M. Cermeño, J. Carro, A. L. Maroto, M. Ángeles Pérez-García,
\textit{Modified Gravity at Astrophysical Scales}, \href{https://iopscience.iop.org/article/10.3847/1538-4357/ab001c}{Astrophysical Journal, {\bf 872} 130, (2019)}. \href{https://arxiv.org/abs/1811.11171}{1811.11171}

\bibitem{SM14} B. F. de Aguiar and R. F. P. Mendes,
\textit{Highly compact neutron stars and screening mechanisms. I. Equilibrium and stability}, \href{https://journals.aps.org/prd/abstract/10.1103/PhysRevD.102.024064}{Phys. Rev. D {\bf 102} 024064, (2020)}. \href{https://arxiv.org/abs/2006.10080}{2006.10080}

\bibitem{Olmo} G.J. Olmo, D. R. Garcia, and A. Wojnar,
\textit{Stellar structure models in modified theories of gravity: lessons and challenges}, \href{https://arxiv.org/abs/1912.05202}{1912.05202}

\bibitem{Eksi} K. Y. Eksi, C. Güngör and M. M. Türkoğlu, 
\textit{What does a measurement of mass and/or radius of a neutron star constrain: Equation of state or gravity?
}, \href{https://journals.aps.org/prd/abstract/10.1103/PhysRevD.89.063003}{PRD {\bf 89} 063003 (2014)}. \href{https://arxiv.org/abs/1402.0488}{1402.0488}

\bibitem{He} X. T. He, F. J. Fattoyev, B. A. Li, and W. G. Newton,
\textit{Impact of the equation-of-state -- gravity degeneracy on constraining the nuclear symmetry energy from astrophysical observables}, \href{https://journals.aps.org/prc/abstract/10.1103/PhysRevC.91.015810}{Phys. Rev. C {\bf 91}, 015810 (2015)}. \href{https://arxiv.org/abs/1408.0857}{1408.0857}

\bibitem{Sakstein} J. Sakstein, 
\textit{Tests of Gravity with Future Space-Based Experiments}, \href{https://journals.aps.org/prd/abstract/10.1103/PhysRevD.97.064028}{Phys. Rev. D \textbf{97}, 064028 (2018)}. \href{https://arxiv.org/abs/1710.03156}{1710.03156}

\bibitem{Mota} D. F. Mota and D. J. Shaw, 
\textit{Strongly Coupled Chameleon Fields: New Horizons in Scalar Field Theory}, \href{https://journals.aps.org/prl/abstract/10.1103/PhysRevLett.97.151102}{Phys. Rev. Lett. \textbf{97} 151102, (2006)}. \href{https://arxiv.org/abs/hep-ph/0606204}{0606204}

\bibitem{ST1983} Shapiro, S.~L. and Teukolsky, S.~A.\ 1983, A Wiley-Interscience Publication, New York: Wiley, 1983.

\bibitem{eos} J. S. Read, B. D. Lackey, B. J. Owen and J. L. Friedman, \textit{Constraints on a phenomenologically parameterized neutron-star equation of state}, \href{https://journals.aps.org/prd/abstract/10.1103/PhysRevD.79.124032}{Phys.Rev.D \textbf{79} 124032, (2009)}. \href{https://arxiv.org/abs/0812.2163}{0812.2163}

\bibitem{eosii} M. F. O'Boyle, C. Markakis, N. Stergioulas and J. S. Read, \textit{A Parametrized Equation of State for Neutron Star Matter with Continuous Sound Speed},  \href{https://arxiv.org/abs/2008.03342}{2008.03342}

\bibitem{2017ApJ...850L..19M} B. Margalit and B. Metzger, \textit{Constraining the Maximum Mass of Neutron Stars from Multi-messenger Observations of GW170817}, The Astrophysical Journal {\bf 850}, L19 (2017).
\href{https://arxiv.org/abs/1710.05938}{1710.05938}

\bibitem{2016MNRAS.459..646B} C. Breu and L. Rezzolla, \textit{Maximum mass, moment of inertia and compactness of relativistic stars}. Monthly Notices of the Royal Astronomical Society {\bf 459}, 646–656 (2016).
\href{https://arxiv.org/abs/1601.06083}{1601.06083}

\bibitem{Jain} B. Jain, V. Vikram and J. Sakstein
\textit{Astrophysical Tests of Modified Gravity: Constraints from Distance Indicators in the Nearby Universe}, \href{https://iopscience.iop.org/article/10.1088/0004-637X/779/1/39}{ Astrophys. J. 779 39 (2013)}. \href{https://arxiv.org/abs/1204.6044}{1204.6044}

\bibitem{Wolf} W. J. Wolf and M. Lagos
\textit{Standard Sirens as a novel probe of dark energy}, \href{https://journals.aps.org/prl/abstract/10.1103/PhysRevLett.124.061101}{ Phys. Rev. Lett. 124, 061101 (2020)}. \href{https://arxiv.org/abs/1910.10580}{1910.10580}

\bibitem{1994ApJ...424..823C} B. Cook, S. Shapiro, and S. Teukolsky, \textit{Rapidly Rotating Neutron Stars in General Relativity: Realistic Equations of State}. The Astrophysical Journal {\bf 424}, 823 (1994)

\bibitem{1994ApJ...422..227C} B. Cook, S. Shapiro, and S. Teukolsky, \textit{Rapidly Rotating Polytropes in General Relativity}. The Astrophysical Journal {\bf 422}, 227 (1994).

\bibitem{ML} C. E. Rasmussen and C. K. I. Williams, \textit{Gaussian Processes for Machine Learning}. The MIT Press (2006).

\bibitem{scikit-learn} F. Pedregosa et al, 
\textit{Scikit-learn: Machine Learning in Python},
\href{https://scikit-learn.org/stable/about.html#citing-scikit-learn}{Journal of Machine Learning Research 12 (2011)}

\bibitem{gamma1} T. Yang, S. Birrer, and B. Hu, 
\textit{The first simultaneous measurement of Hubble constant and post-Newtonian parameter from Time-Delay Strong Lensing}, \href{https://academic.oup.com/mnrasl/article-abstract/497/1/L56/5859500?redirectedFrom=fulltext}{Mon. Not. Roy. Astron. Soc. \textbf{497} 1 (2020)}. \href{https://arxiv.org/abs/2003.03277}{2003.03277}

\bibitem{gamma2} D. Jyoti, J. B. Munoz, R. R. Caldwell, and M. Kamionkowski,
\textit{Cosmic Time Slip: Testing Gravity on Supergalactic Scales with Strong-Lensing Time Delays}, \href{https://journals.aps.org/prd/abstract/10.1103/PhysRevD.100.043031}{Phys. Rev. D \textbf{100}, 043031 (2019)}. \href{https://arxiv.org/abs/1906.06324}{1906.06324}




\end{thebibliography}
\end{document}